\documentclass[aps,prc,superscriptaddress,twoside,twocolumn,nofootinbib,10pt,%
showpacs,floatfix]{revtex4-1}

\usepackage{amsmath,amssymb}
\usepackage{graphicx,bm}
\usepackage{epstopdf}
\usepackage{ulem} 
\usepackage[usenames]{color}
\usepackage{float}

\allowdisplaybreaks

\begin{document}

\title{Coupled channel analysis of molecule picture of $P_{c}(4380)$}

\author{Yuki Shimizu}
\email{yshimizu@hken.phys.nagoya-u.ac.jp}
\affiliation{Department of Physics,  Nagoya University, Nagoya, 464-8602, Japan}

\author{Daiki Suenaga}
\email{suenaga@hken.phys.nagoya-u.ac.jp}
\affiliation{Department of Physics,  Nagoya University, Nagoya, 464-8602, Japan}

\author{Masayasu Harada}
\email{harada@hken.phys.nagoya-u.ac.jp}
\affiliation{Department of Physics,  Nagoya University, Nagoya, 464-8602, Japan}

\date{\today}


\begin{abstract}
We construct a potential obtained by one-pion exchange for the coupled channel $\Sigma_c^\ast$$\bar{D}$-$\Sigma_c$$\bar{D}^\ast$, and solve the coupled Schr\"{o}dinger equations to determine the binding energy.
We find that there exists one or two bound states with the binding energy of several MeV  below the threshold of $\Sigma_c^\ast$ and $\bar{D}$, dominantly made from a $\Sigma_c^\ast$ baryon and a $\bar{D}$ meson, with the size of about $1.5$\,fm for a wide parameter region.  
We also study the pentaquark states including a $b$ quark and/or an anti-$b$ quark.
We show that there exist pentaquarks including $c\bar{b}$, $b\bar{c}$, and $b\bar{b}$, all of which lie at about $10$\,MeV below the corresponding threshold and have size of about $1.5$\,fm.
\end{abstract}

\maketitle


\section{Introduction}

Hadrons made of more than three quarks are interesting objects to study.
In the summer of 2015, the LHCb announced the discovery  of the hidden charm pentaquark~\cite{Aaij:2015tga}: 
one has a mass of $4380\pm8\pm29$\,MeV and a width of $205\pm18\pm86$\,MeV, while the second is narrower, with a mass of $4449.8\pm1.7\pm2.5$\,MeV and a width of $39\pm5\pm19$\,MeV. 
Soon after the announcement, there appeared many theoretical analyses on the pentaquark based on the molecular picture~
\cite{Chen:2015loa,He:2015cea,Chen:2015moa,Karliner:2015ina,Huang:2015uda,Roca:2015dva,Mironov:2015ica,Meissner:2015mza,Xiao:2015fia,Burns:2015dwa,Wang:2015qlf,Kahana:2015tkb,Chen:2016heh,Chen:2016otp,Lu:2016nnt},
the rescattering effects~\cite{Guo:2015umn,Liu:2015fea,Mikhasenko:2015vca,Liu:2016dli},
the diquark-diquark-antiquark (or diquark-triquark) picture
~\cite{Maiani:2015vwa,Lebed:2015tna,Anisovich:2015cia,Li:2015gta,Anisovich:2015zqa,Wang:2015wsa,Ghosh:2015ksa,Maiani:2015iaa,Wang:2015epa,Wang:2015ava,Zhu:2015bba,Wang:2015ixb}, and so on~\cite{Wang:2015jsa,Kubarovsky:2015aaa,Scoccola:2015nia,Hsiao:2015nna,Karliner:2015voa,Aaij:2015fea,Anisovich:2015xja,Stone:2015iba,Cheng:2015cca,Lu:2015fva,Yang:2015bmv,Wang:2015pcn,Gerasyuta:2015djk,Schmidt:2016cmd,Roca:2016tdh}, in addition to some relevant  works~\cite{Wu:2010jy,Wu:2010vk,Wang:2011rga,Yang:2011wz,Wu:2012md,Xiao:2013yca} done before the LHCb result.

There are many analyses for the molecule picture.
In Ref.~\cite{Chen:2015loa}, the pentaquarks are regarded as the bound states of the $\bar{D}^\ast$ meson and the  $\Sigma_c$  baryon by using the potential made by the 
 one-pion exchange. 
The contributions from the $\sigma$ and $\omega$ mesons are further included in the potential~\cite{He:2015cea}, which shows that $P_c(4380)$ can be understood as a bound state of $\Sigma_c^\ast$ and $\bar{D}$.
In Ref.~\cite{Chen:2015moa}, the QCD sum rule is used to show that $P_c(4380)$ is a bound state of $\Sigma_c$ and $\bar{D}^\ast$, and that $P_c(4450)$ is a bound state of a mixture of $\Lambda_c$$\bar{D}^\ast$ and $\Sigma_c^\ast$$\bar{D}$.
An analysis based on a quark model was 
performed~\cite{Wang:2011rga}
before the LHCb result, which showed that there exists a bound state of $\Sigma_c$ and $\bar{D}$ with the threshold being about 4.3\,GeV.
There are many other analyses such as those in Refs.~\cite{Karliner:2015ina,Huang:2015uda,Roca:2015dva,Meissner:2015mza,Xiao:2015fia,Burns:2015dwa,Wang:2015qlf,Chen:2016heh,Chen:2016otp} showing several different molecule structures.

The recently observed $P_c(4380)$ lies below the $\Sigma_c^*$$\bar{D}$ threshold in several MeV, so that this new state can be naturally regarded as a molecular state of $\Sigma_c^*$$\bar{D}$. However, it is impossible to construct a $\Sigma_c^\ast$$\bar{D}$  molecular  state by a potential made by just one-pion exchange  because $\bar{D}\bar{D}\pi$ vertex is prohibited by the parity invariance. Then, we need to take into account effects of  coupled channels to study the existence of the molecular state mainly made from $\Sigma_c^\ast\bar{D}$ by the one-pion exchange. 
The most likely channel coupled to $\Sigma_c^{\ast}\bar{D}$ through the one-pion exchange is the $\Sigma_c\bar{D}^{\ast}$ channel, since sum of their masses is closer to the sum of masses of $\Sigma_c^{\ast}$ and $\bar{D}$ than the other channels.
Thus in this paper, we investigate the coupled channel effect of $\Sigma_c^\ast$$\bar{D}$-$\Sigma_c$$\bar{D}^\ast$ to molecular states.
As pointed out in Ref.~\cite{Chen:2016qju}, this coupled channel effect was not yet studied.
In the present analysis, we construct a one-pion exchange potential 
following the procedure explained in Ref.~\cite{Yamaguchi:2014era} and solve the Schr\"odinger type equation of motion.
Our results show that the binding energy of the ground state is about several MeV below the sum of $\Sigma_c^\ast$ and $\bar{D}$ masses of $4385.3$\,MeV in the wide range of the relevant parameters, and that the percentage of the $\Sigma_c^\ast$$\bar{D}$ component is more than 99\%.
This implies that the observed $P_c(4380)$ can be reasonably understood as a molecular state dominantly made from the $\Sigma_c^\ast$ baryon and the $\bar{D}$ meson.

This paper is organized as follows.
In Sec.~\ref{sec:opep} we construct a potential by one-pion exchange.
Then, we make a numerical analysis in Sec.~\ref{sec:results}.
We extend the analysis by replacing the charm quark with the bottom quark in Sec.~\ref{sec:bottom}.
Finally, a summary and discussions are given in Sec.~\ref{sec:Summary}.


\section{One-pion exchange potential for $\Sigma_{c}^{*} \bar{D}$-$\Sigma_{c} \bar{D}^{*}$ channels}
\label{sec:opep}

In this section, we construct a potential for $\Sigma_{c}^{*} \bar{D}$-$\Sigma_{c} \bar{D}^{*}$ channels generated by one-pion exchange.

Here, we first specify interactions of relevant hadrons with the pions based on the heavy quark symmetry and the chiral symmetry.
The pion field is introduced into our model within the framework of the chiral Lagrangian  based on the spontaneous chiral symmetry breaking of $\mbox{SU}(2)_{\rm R}\times\mbox{SU}(2)_{\rm L} \to \mbox{SU}(2)_{\rm V}$.
The basic quantity is 
\begin{equation}
\alpha_{\perp \mu} = \frac{1}{2i} \left[ \partial_\mu \xi \cdot \xi^\dag - \partial_\mu \xi^\dag \cdot \xi \right]
\ ,
\end{equation}
where $\xi = e^{i \pi / f_\pi}$ with $\pi = \pi_a T_a$ ($a=1,2,3$) and $f_\pi=92.4\,\mbox{MeV}$ being the pion fields and the pion decay constant. 
The quantity $\alpha_{\perp \mu}$ transforms as 
\begin{equation}
\alpha_{\perp\mu} \ \to \ h \, \alpha_{\perp\mu} \, h^\dag \ ,
\end{equation}
where $h$ is an element of $\mbox{SU}(2)_{\rm V}$.

We include the $\bar{D}$ and $\bar{D}^\ast$ fields through the standard heavy meson effective field expressed as
\begin{equation}
\bar{H} = \left[ \bar{D}^{\ast \mu}\gamma_{\mu} - \bar{D}\gamma_{5} \right] \frac{1+v\hspace{-.47em}/}{2} \ ,
\end{equation}
where $v^\mu$ denotes the velocity of the heavy meson, $\bar{D}$ and $\bar{D}^\ast$ are the isodoublet fields for the fluctuation of the heavy mesons, $\bar{D}^{+,0}$ and $\bar{D}^{\ast+,0}$. 
Under the chiral transformation, $\bar{H}$ transforms as
\begin{equation}
\bar{H} \ \to \ h \, \bar{H} \ .
\end{equation}
By using this together with $\alpha_{\perp\mu}$ for the pion fields,
an interaction for heavy mesons with pions with least derivatives is written as~\cite{Wise:1992hn,Yan:1992gz,Cho:1992gg}
\begin{align}
\mathcal{L}_{int} = g\textrm{Tr}\left[ H\gamma_{\mu}\gamma_{5} \alpha_{\perp}^{\mu} \bar{H} \right] \ ,
\label{L int H}
\end{align}
where $g$ is a dimensionless coupling constant.
Expanding the $H$ fields and $\alpha_{\perp\mu}$, the one-pion interaction terms of the heavy mesons are expressed as
\begin{align}
\mathcal{L}_{int} = &\left( \frac{2g}{f_{\pi}}\bar{D}_{\mu}^{\ast \dagger}\partial^{\mu}\pi\bar{D} + \mbox{h.c.} \right) + \frac{2ig}{f_{\pi}}\epsilon^{\mu\nu\rho\sigma}v_{\mu}\bar{D}_{\nu}^{\ast \dagger}\partial_{\rho}\pi \bar{D}_{\sigma}^{\ast}\ .
\end{align}

The relevant baryons $\Sigma_c$ and $\Sigma_c^\ast$ are included through an isotriplet heavy-quark doublet field $S_\mu$ as
\begin{align}
S_{\mu} = -\sqrt{\frac{1}{3}}\left( \gamma_{\mu} + v_{\mu} \right) \gamma_{5}\Sigma_c + \Sigma_{c\,\mu}^{\ast} \ .
\end{align}
These two fields are expressed in the isospin space as
\begin{align}
\Sigma_c= \left(
\begin{array}{cc}
\Sigma_{c}^{++} & \frac{\Sigma_{c}^{+}}{\sqrt{2}} \\
\frac{\Sigma_{c}^{+}}{\sqrt{2}} & \Sigma_{c}^{0} \\
\end{array}
\right) \ , \quad \Sigma^{\ast}_{c\,\mu} = \left(
\begin{array}{cc}
\Sigma_{c}^{\ast ++} & \frac{\Sigma_{c}^{\ast +}}{\sqrt{2}} \\
\frac{\Sigma_{c}^{\ast +}}{\sqrt{2}} & \Sigma_{c}^{\ast 0} \\
\end{array}
\right)_\mu \ .
\end{align}
The $S_{\mu}$ field transforms under the $\mbox{SU}(2)_R\times \mbox{SU}(2)_L$ chiral transformation as 
\begin{equation}
S_{\mu} \ \to \ h \, S_{\mu} \, h^{T} \ .
\end{equation}
An interaction Lagrangian with least derivative is expressed as~\cite{Yan:1992gz,Cho:1992gg,Liu:2011xc}
\begin{align}
\mathcal{L}_{int} = -\frac{3}{2}i g_{1}\epsilon^{\mu\nu\rho\sigma}v_{\sigma} \textrm{Tr}\left[ \bar{S}_{\mu}\alpha_{\perp\nu}S_{\rho} \right] \ ,
\label{L int B}
\end{align}
which leads to the following one-pion interaction terms:
\begin{align}
\mathcal{L}_{int} = &\frac{ig_{1}}{2f_{\pi}}\epsilon^{\mu\nu\rho\sigma}v_{\sigma}\textrm{Tr}\left[ \bar{\Sigma}_c \gamma_{\mu}\gamma_{\rho}\partial_{\nu}\pi \Sigma_c \right] \nonumber \\
&-\frac{3ig_{1}}{2f_{\pi}}\epsilon^{\mu\nu\rho\sigma}v_{\sigma}\textrm{Tr}\left[ \bar{\Sigma}_{c\,\mu}^{\ast}\partial_{\nu}\pi \Sigma_{c\,\rho}^{\ast} \right] \nonumber \\
&+\left( \frac{\sqrt{3}ig_{1}}{2f_{\pi}}\epsilon^{\mu\nu\rho\sigma}v_{\sigma}\textrm{Tr} \left[ \bar{\Sigma}_{c\,\mu}^{\ast}\partial_{\nu}\pi\gamma_{\rho}\gamma_{5}\Sigma_c \right] + H.c. \right) .
\end{align}

We construct a one-pion exchange potential (OPEP) between ($\bar{D}$, $\bar{D}^\ast$) mesons and ($\Sigma_c$, $\Sigma_c^\ast$) baryons from the above interaction terms.
Following the procedure explained in Ref.~\cite{Yamaguchi:2014era},
we introduce the monopole-type form factor at each vertex given by
\begin{align}
F(\vec{q}) = \frac{\Lambda^{2} - m_{\pi}^{2}}{\Lambda^{2}+|\vec{q}|^2}
\end{align}
where $m_{\pi}$ is the pion mass, $\vec{q}$ is the momentum of the pion, and $\Lambda$ is  a cutoff parameter. Although the cutoff $\Lambda$ for the meson-pion vertex may not be the same as that for the baryon-pion vertex, we use the same parameter in the present analysis for simplicity. 
By including this form factor, the OPEPs
for the S-wave channels of $\Sigma_{c}^{\ast}\bar{D}$-$\Sigma_{c}^{\ast}\bar{D}$, $\Sigma_{c}\bar{D}^{\ast}$-$\Sigma_{c}\bar{D}^{\ast}$ and $\Sigma_{c}^{\ast}\bar{D}$-$\Sigma_{c}\bar{D}^{\ast}$ with $I(J^{P})=\frac{1}{2}(\frac{3}{2}^{-})$ are obtained as
\begin{align}
V_{\Sigma_{c}^{\ast}\bar{D}-\Sigma_{c}^{\ast}\bar{D}}(r)&=0 \\
V_{\Sigma_{c}\bar{D}^{\ast}-\Sigma_{c}\bar{D}^{\ast}}(r)&=-\frac{1}{3}\times\frac{g_{1}gm_{\pi}^3}{8\pi f_{\pi}^2}Y_{1}(m_{\pi}, \Lambda, r) \\
V_{\Sigma_{c}\bar{D}^{\ast}-\Sigma_{c}^{\ast}\bar{D}}(r)&=-\frac{1}{2\sqrt{3}}\times\frac{g_{1}gm_{\pi}^3}{8\pi f_{\pi}^2}Y_{1}(m_{\pi}, \Lambda, r)
\ ,
\end{align}
where 
$Y_{1}(m_{\pi}, \Lambda, r)$ is defined as
\begin{align}
Y_{1}(m_{\pi}, \Lambda, r) = Y(m_{\pi}r) - \frac{\Lambda}{m_{\pi}}Y(\Lambda r) - \frac{\Lambda^2-m_{\pi}^2}{2m_{\pi}\Lambda}e^{-\Lambda r}, 
\end{align}
with $Y(x)=\frac{e^{-x}}{x}$. 
It should be noted that the OPEP for the $\Sigma_{c}^{\ast}\bar{D}$-$\Sigma_{c}^{\ast}\bar{D}$ channel is zero because the $\bar{D}\bar{D}\pi$ vertex vanishes by parity. 


\section{Numerical results for the binding energy and the mixing structure}
\label{sec:results}

The relevant Schr\"{o}dinger equation is expressed as
\begin{equation}
\left[ - \frac{1}{2m} \vec{\nabla}^2 + V(r) \right] \Psi\left(\vec{r}\right) = E \Psi\left(\vec{r}\right) \ ,
\end{equation}
where $m$ is the reduced mass, $E$ is the energy eigenvalue, 
 $V(r)$ is the potential matrix obtained from the OPEPs in the previous section as
\begin{align}
V(r) & = \begin{pmatrix} 
V_{\Sigma_{c}^{\ast}\bar{D}-\Sigma_{c}^{\ast}\bar{D}}(r) & 
  V_{\Sigma_{c}\bar{D}^{\ast}-\Sigma_{c}^{\ast}\bar{D}}(r) \\
V_{\Sigma_{c}\bar{D}^{\ast}-\Sigma_{c}^{\ast}\bar{D}}(r) & 
  V_{\Sigma_{c}\bar{D}^{\ast}-\Sigma_{c}\bar{D}^{\ast}}(r) \\
\end{pmatrix}
\\
&= \begin{pmatrix} 0 & - \frac{1}{2\sqrt{3}} \\ - \frac{1}{2\sqrt{3}} & - \frac{1}{3} \end{pmatrix} \times \frac{g_{1}gm_{\pi}^3}{8\pi f_{\pi}^2}Y_{1}(m_{\pi}, \Lambda, r)
\ .
\end{align}
The wave function $\Psi\left(\vec{r}\right)$ has two components for the $\Sigma_{c}^{\ast}\bar{D}$ and $\Sigma_{c}\bar{D}^{\ast}$ states:
\begin{equation}
\Psi = \begin{pmatrix} \psi_{\Sigma_{c}^{\ast}\bar{D}} \\ \psi_{\Sigma_{c}\bar{D}^{\ast}} \\ \end{pmatrix} \ .
\end{equation}
Solving the above Schr\"{o}dinger equation, we determine the binding energy of the bound states and the mixing structure. 
We use $m_{\pi}$=137.2MeV, $m_{\Sigma_{c}}$=2453.5MeV, $m_{\Sigma_{c}^{\ast}}$=2518.1MeV, $m_{\bar{D}}$=1867.2MeV, $m_{\bar{D}^{\ast}}$=2008.6MeV for the hadron masses. 
For the coupling constant among one pion and the charmed mesons $g$ defined in Eq.~(\ref{L int H}), we use $\vert g \vert=0.60$ determined from the $D^\ast \to D \pi$ decay width~\cite{Agashe:2014kda}.
For the one-pion coupling of charmed baryons $g_1$ defined in Eq.~(\ref{L int B}),  we take $g_1= 0.95$ as an example which is close to the value $0.94$ estimated in a quark model~\cite{Liu:2011xc}, and study the dependence by using $g_1 = 0.75$ and $1.95$.
We also vary the value of the cutoff parameter $\Lambda$ for the form factor from $0.8$\,GeV to $2.5$\,GeV.

\begin{figure}[!htbp]
\begin{center}
(a) \\
\includegraphics[width=0.45\textwidth, bb=0 0 609 321, clip]{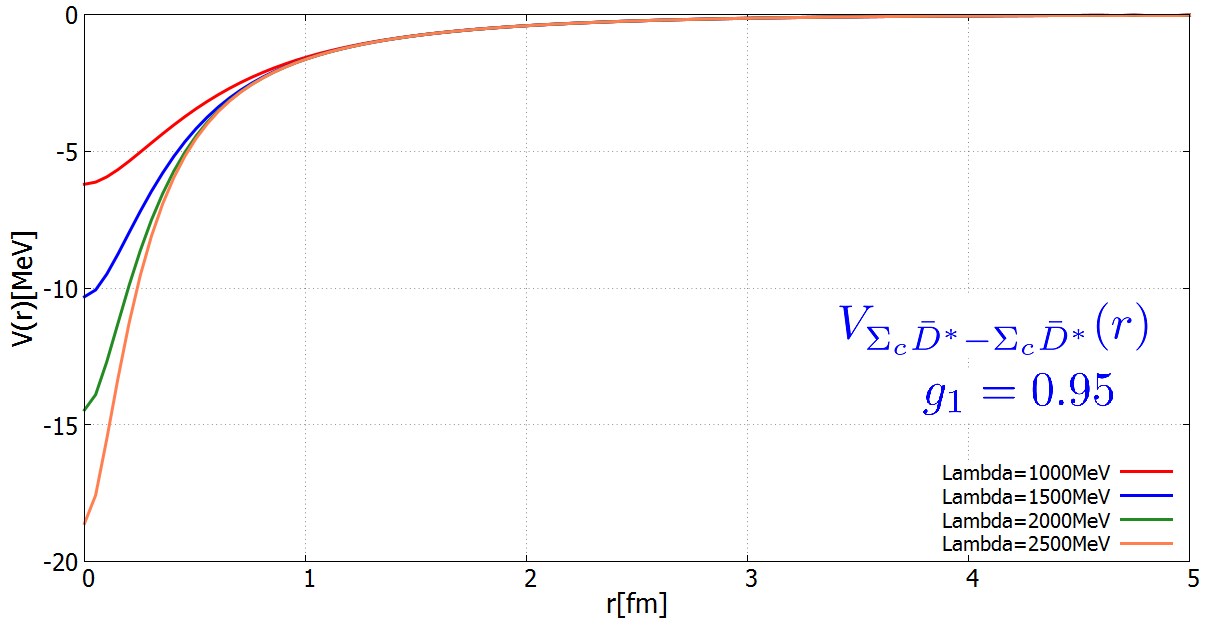}\\
(b) \\
\includegraphics[width=0.45\textwidth, bb=0 0 609 321, clip]{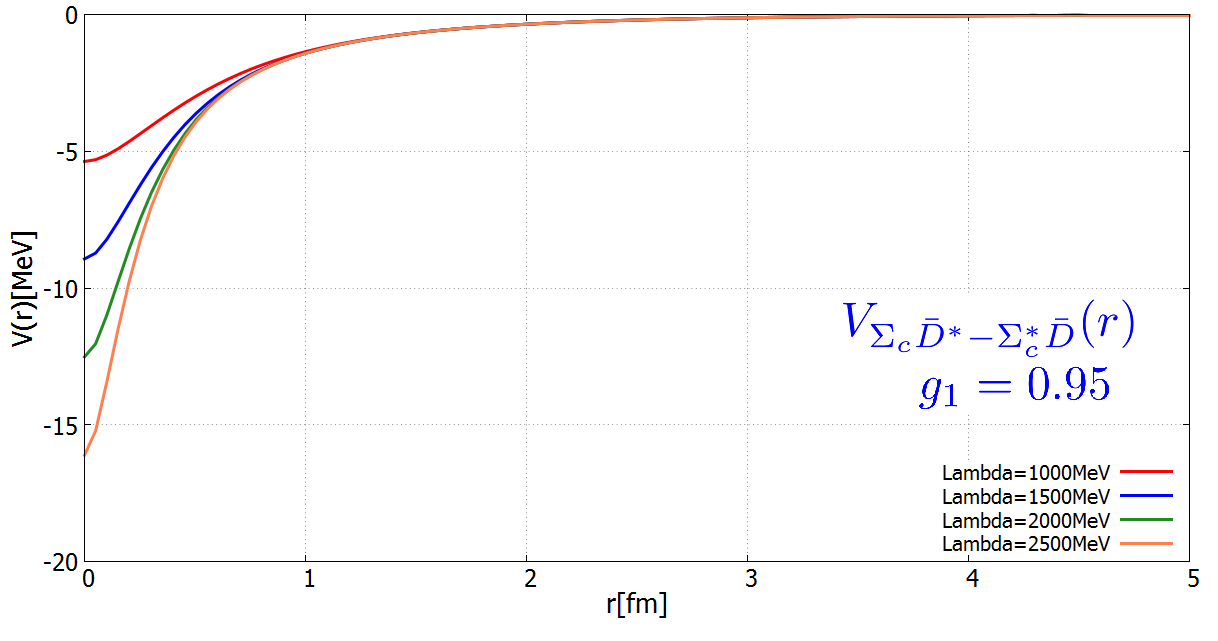}
\caption{(color online) 
One-pion exchange potentials (a) $V_{\Sigma_c\bar{D}^{\ast}-\Sigma_c\bar{D}^{\ast}}$ and (b) $V_{\Sigma_c\bar{D}^{\ast}-\Sigma_c^{\ast}\bar{D}}$ for several choices of the cutoff parameter $\Lambda$ with fixed value of $g_1$=0.95.
}
\label{fig:OPEP}
\end{center}
\end{figure}

We first show $r$ dependences of two potentials 
$V_{\Sigma_{c}\bar{D}^{\ast}-\Sigma_{c}^{\ast}\bar{D}}(r)$ and $V_{\Sigma_{c}\bar{D}^{\ast}-\Sigma_{c}\bar{D}^{\ast}}(r)$
in Fig.~\ref{fig:OPEP} for 
several choices of the cutoff parameter $\Lambda$ with fixed value of 
$g_1=0.95$ as an example. 
We note that the shape of the potential for $\Sigma_c \bar{D}^{\ast}$-$\Sigma_c \bar{D}^{\ast}$ shown in Fig.\ref{fig:OPEP}(a) is different from the one shown in Ref.\cite{Chen:2015loa}. This may be since our regularization method following Ref.\cite{Yamaguchi:2014era} is different from the one adopted in Ref.\cite{Chen:2015loa} .

\begin{figure}[!htbp]
\begin{center}
\includegraphics[width=0.45\textwidth, bb=0 0 1218 641, clip]{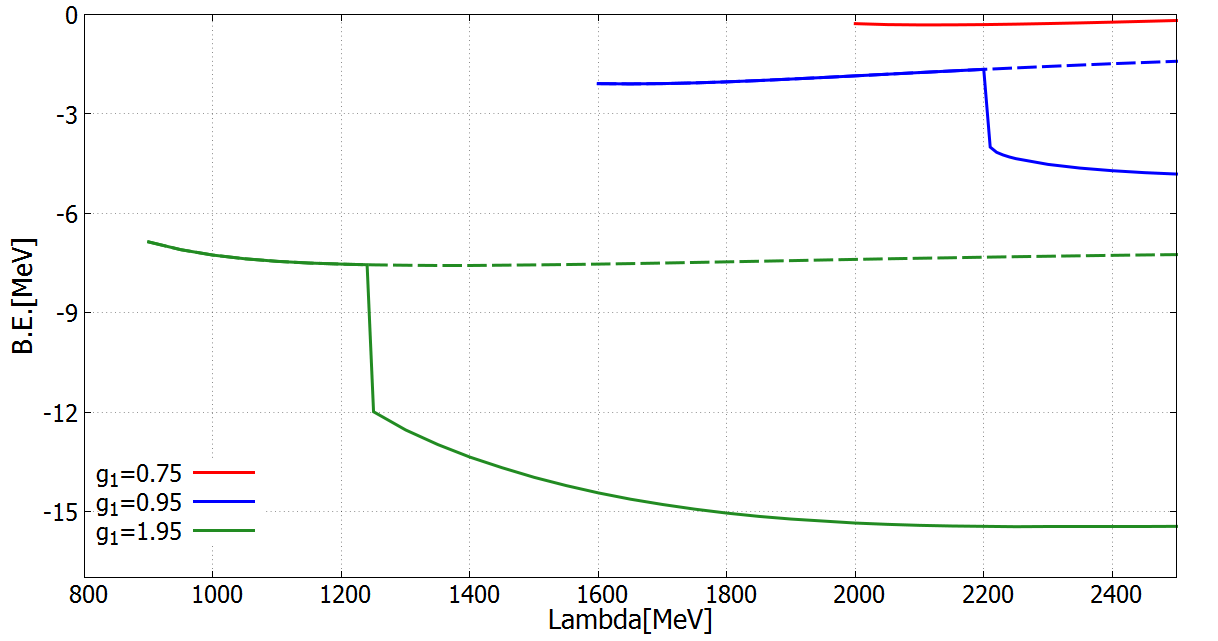}
\caption{(color online) 
Binding energy(B.E.) 
for $\Sigma_{c}\bar{D}^{\ast}$-$\Sigma_{c}^{\ast}\bar{D}$ molecular state measured from the $\Sigma_{c}^{\ast} \bar{D}$ threshold of  $4385.3$\,MeV plotted against the cutoff $\Lambda$ for the form factor.
The values of B.E. for the ground states are shown by solid curves and those for the first excited states are by dashed curves.
The red, blue and green curves are for $g_1 = 0.75$, $0.95$, and $1.95$, respectively. 
}
\label{fig:c-barc}
\end{center}
\end{figure}

\begin{figure}[!tbp]
\begin{center}
\includegraphics[width=0.45\textwidth, bb=0 0 1218 641, clip]{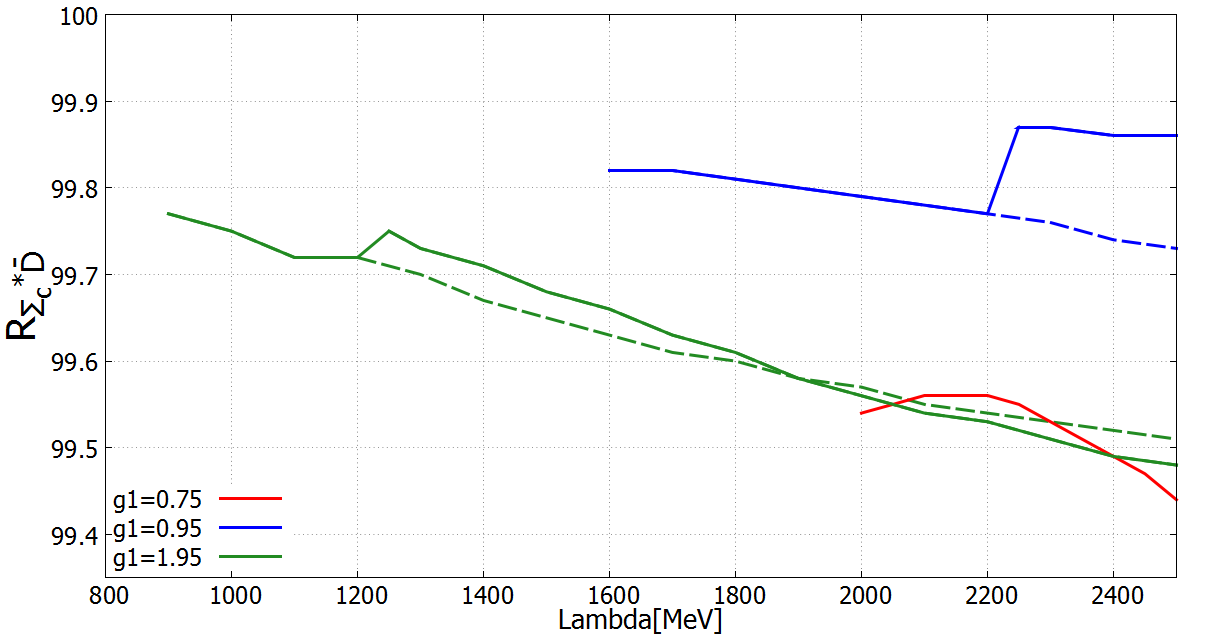}
\caption{(color online) Percentage of the $\Sigma_{c}^{\ast}\bar{D}$ component of the wave function for the ground state defined in Eq.~(\ref{eq:percentage}), plotted against the cutoff $\Lambda$.
The values of the percentage for ground states are shown by solid curves and those for the first excited states are by dashed curves.
The red, blue and green curves are for $g_1 = 0.75$, $0.95$, and $1.95$, respectively.
}
\label{fig:c-barc-ratio}
\end{center}
\end{figure}

\begin{figure}[!htbp]
\begin{center}
\includegraphics[width=0.45\textwidth, bb=0 0 1218 641, clip]{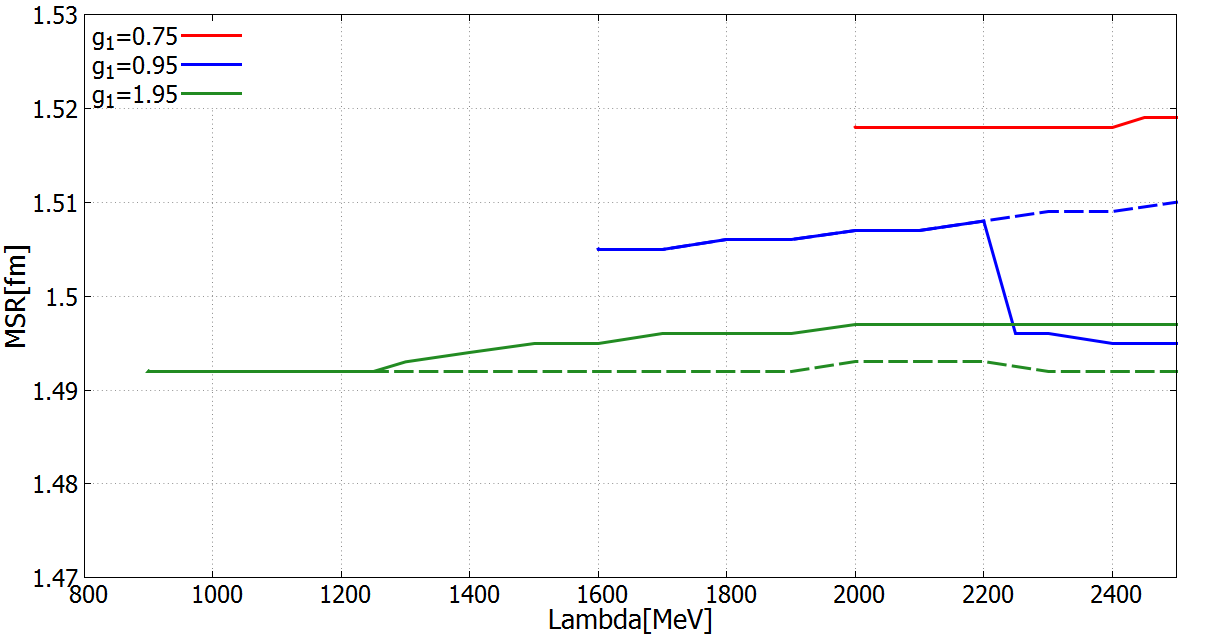}
\caption{(color online) Mean square radius (MSR) for $\Sigma_c\bar{D}^{\ast}$-$\Sigma_c^{\ast}\bar{D}$ system, plotted against the cutoff $\Lambda$.
The values of the percentage for ground states are shown by solid curves and those for the first excited states are by dashed curves.
The red, blue and green curves are for $g_1 = 0.75$, $0.95$, and $1.95$, respectively.
}
\label{fig:c-barc-MSR}
\end{center}
\end{figure}

\begin{figure}[!htbp]
\begin{center}
\includegraphics[width=0.45\textwidth, bb=0 0 1218 641, clip]{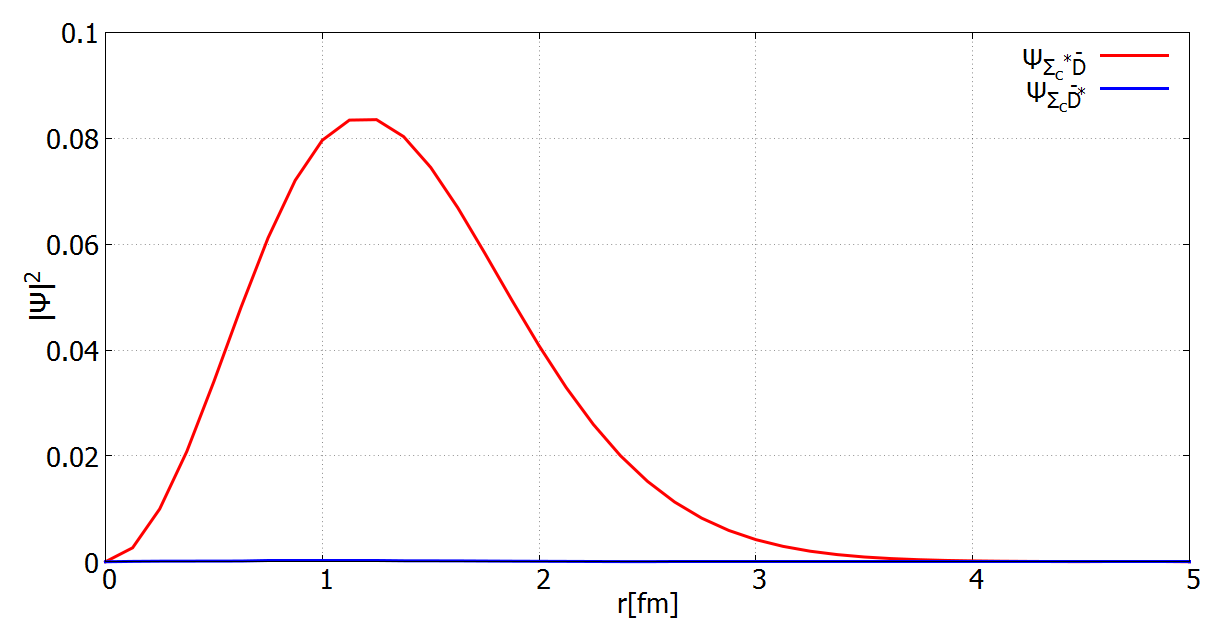}
\caption{(color online) $r$ dependences of the squared wave functions for $\Sigma_c^\ast\bar{D}$ and $\Sigma_c\bar{D}^\ast$ components with fixed values of $\Lambda=1600$\,MeV and $g_1=0.95$.
}
\label{fig:wavefunction}
\end{center}
\end{figure}

Next, we plot 
the resultant values of the binding energy against the cutoff parameter $\Lambda$ for fixed values of $g_1=0.75,0.95$ and $1.95$ in Fig.~\ref{fig:c-barc}. In this plot, we measure the binding energy from the $\Sigma_{c}^{\ast} \bar{D}$ threshold of $4385.3$\,MeV.
For studying the mixing structure of these bound states, 
we plot the percentage of the $\Sigma_c^\ast\bar{D}$ component of the wave function defined as
\begin{equation}
R_{\Sigma_c^{\ast}\bar{D}}=
\frac{ \displaystyle \int d^3r \left \vert \psi_{\Sigma_c^\ast\bar{D}}\left(\vec{r}\right) \right\vert^2 }
{ \displaystyle \int d^3r \left[ \left \vert \psi_{\Sigma_c^\ast\bar{D}}\left(\vec{r}\right) \right\vert^2 
  + \left \vert \psi_{\Sigma_c\bar{D}^\ast}\left(\vec{r}\right) \right\vert^2 \right] }
\label{eq:percentage}
\end{equation}
in Fig.~\ref{fig:c-barc-ratio}.
To see the size of the bound states,
we show the mean square radius (MSR) for the bound states defined by $\sqrt{ \left\langle r^2 \right\rangle }$ , 
where
\begin{equation}
\left\langle r^2 \right\rangle = 
\frac
{ \displaystyle \int d^3r \, \vec{r}\, ^2 \left[ \left \vert \psi_{\Sigma_c^\ast\bar{D}}\left(\vec{r}\right) \right\vert^2 
  + \left \vert \psi_{\Sigma_c\bar{D}^\ast}\left(\vec{r}\right) \right\vert^2 \right] }
{ \displaystyle \int d^3r \left[ \left \vert \psi_{\Sigma_c^\ast\bar{D}}\left(\vec{r}\right) \right\vert^2 
  + \left \vert \psi_{\Sigma_c\bar{D}^\ast}\left(\vec{r}\right) \right\vert^2 \right] }
\end{equation}
in Fig.~\ref{fig:c-barc-MSR}.
We also plot the $r$ dependence of the wave functions of the $\Sigma_c^\ast\bar{D}$ and $\Sigma_c\bar{D}^\ast$ component with the fixed values of $\Lambda=1600$\,MeV and $g_1=0.95$ in Fig.~\ref{fig:wavefunction}.

From Figs.~\ref{fig:c-barc}-\ref{fig:c-barc-MSR} together with Fig.~\ref{fig:wavefunction}, 
we can see the following properties:
There are bound states with the binding energy of several MeV dominantly (more than $99$\%) made from $\Sigma_c^\ast\bar{D}$ with the size of about $1.5$\,fm in wide parameter range. 
Inside a bound state, the distance between the $\Sigma_c^\ast$ and $\bar{D}$ components is about $1$\,fm, which implies that it is naturally regarded as a molecule state.
The binding energy and the MSR are rather stable against the change of $\Lambda$ in most regions, while the percentage slightly decreases with increasing $\Lambda$.
When the value of $\Lambda$ is increased with a fixed value of $g_1$,  three quantities of the ground state shown by solid curves suddenly change their values at a certain cutoff, e.g., at $\Lambda=2200$\,MeV for $g_1=0.95$, the binding energy jumps from $E\sim 1.5$\,MeV to $4$\,MeV.   
But the values before the jump are smoothly connected to those of the first excited states drawn by dashed curves.  As a result, there are two bound states for the large values of the cutoff $\Lambda$ and/or the coupling $g_1$.
We can understand these properties as follows:
The binding energy and the size (MSR) are determined by the shape of the potential and the kinetic energy.
When the cutoff $\Lambda$ is increased, the shape of the potential is changed, i.e., the depth becomes deep. 
On the other hand, the kinetic energy by the quantum fluctuation is stable since the reduced mass is unchanged.  
Therefore, when the $\Lambda$ reaches a certain value, the potential energy exceeds the value for which the first excited state can exist. Then, there is a jump of three quantities.

From the above analysis, we conclude that there are one or two bound states in the coupled channel of $\Sigma_c^\ast\bar{D}$ and $\Sigma_c\bar{D}^\ast$ with the binding energy of several MeV and the size of about $1.5$\,fm dominantly made from a $\Sigma_c^\ast$ baryon and a $\bar{D}$ meson.
Since the sum of the masses of $\Sigma_c^\ast$ and $\bar{D}$ is $4385.3$\,MeV, and the observed mass of $P_{c}(4380)$ is $4380\pm 8 \pm 29 $\,MeV, then the obtained binding energy is just suitable for considering $P_{c}(4380)$ as a molecular state existing in the coupled channel of $\Sigma_{c}$$\bar{D}^{\ast}$-$\Sigma_{c}^{\ast}$$\bar{D}$.
Furthermore, for some parameter region, there exist two molecular states within a few MeV range.


\section{Pentaquarks including a $b$ quark and/or a $\bar{b}$ quark}
\label{sec:bottom}

In this section, we extend our analysis in the previous section to pentaquarks 
including a $b$ quark and/or a $\bar{b}$ quark.
As in the case of the charmed baryons and mesons, we use the heavy-quark spin symmetry to relate the $B^\ast$$B$$\pi$ coupling to $B^\ast$$B^\ast$$\pi$ coupling as well as the $\Sigma_{b}^\ast$$\Sigma_b$$\pi$ coupling to the $\Sigma_b^\ast$$\Sigma_b^\ast$$\pi$ coupling.
The heavy-quark flavor symmetry further relates these couplings to the ones for the charmed hadrons.
Then, in the present analysis, we fix $|g_{B^\ast B \pi}| = |g_{B^\ast B^\ast \pi}| = |g_{D^\ast D \pi}| = |g_{D^\ast D^\ast \pi}| = 0.60$ and vary the value of $g_{\Sigma_{b}^\ast \Sigma_b \pi} = g_{\Sigma_b^\ast \Sigma_b^\ast \pi}$ from $0.75$ to $1.95$.
As in the previous section, we introduce one common cutoff parameter $\Lambda$ for two form factors, and study the dependence of the results.

\begin{figure}[!htbp]
\begin{center}
(a) \\
\includegraphics[width=0.45\textwidth, bb=0 0 1218 641, clip]{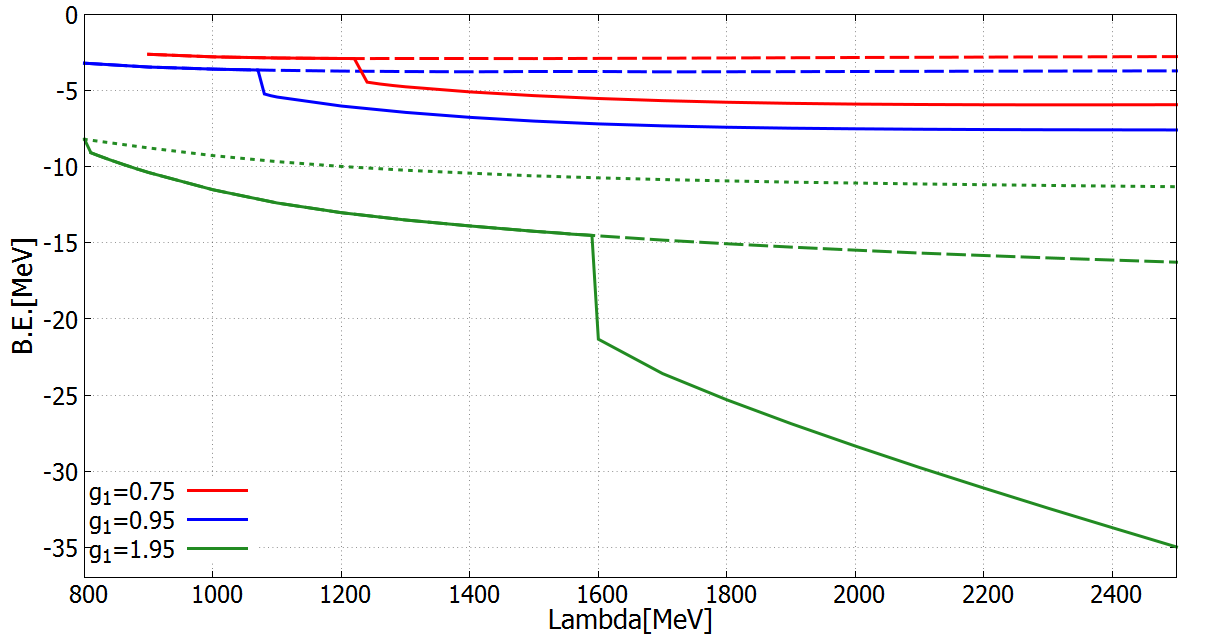} \\
(b) \\
\includegraphics[width=0.45\textwidth, bb=0 0 1218 641, clip]{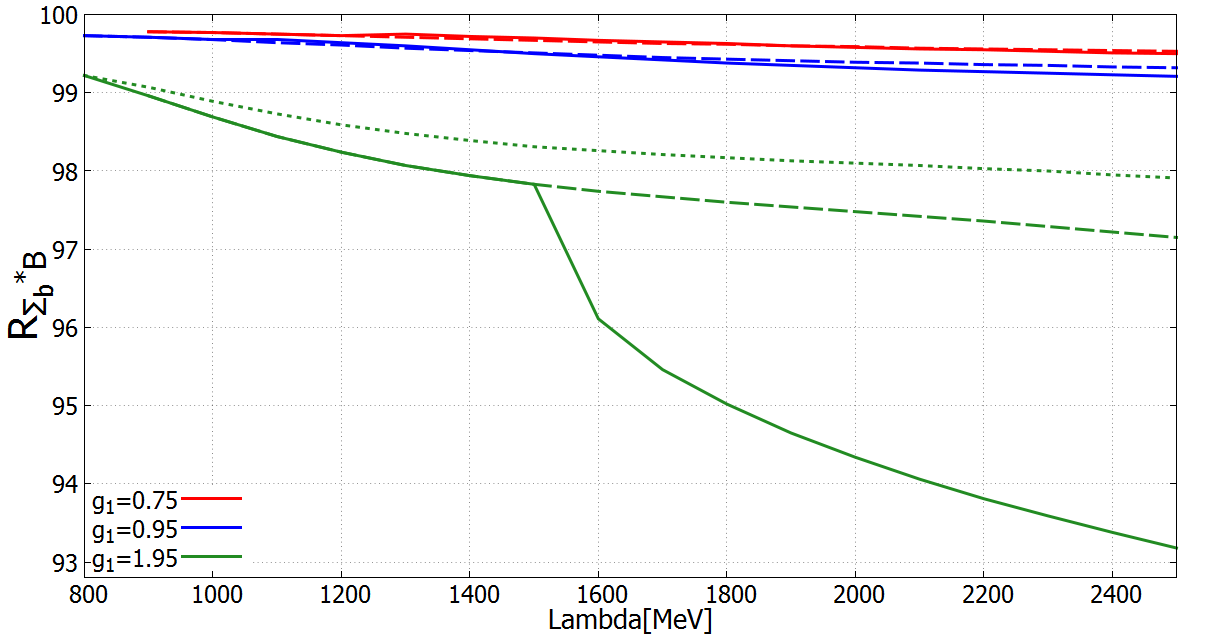} \\
(c) \\
\includegraphics[width=0.45\textwidth, bb=0 0 1218 641, clip]{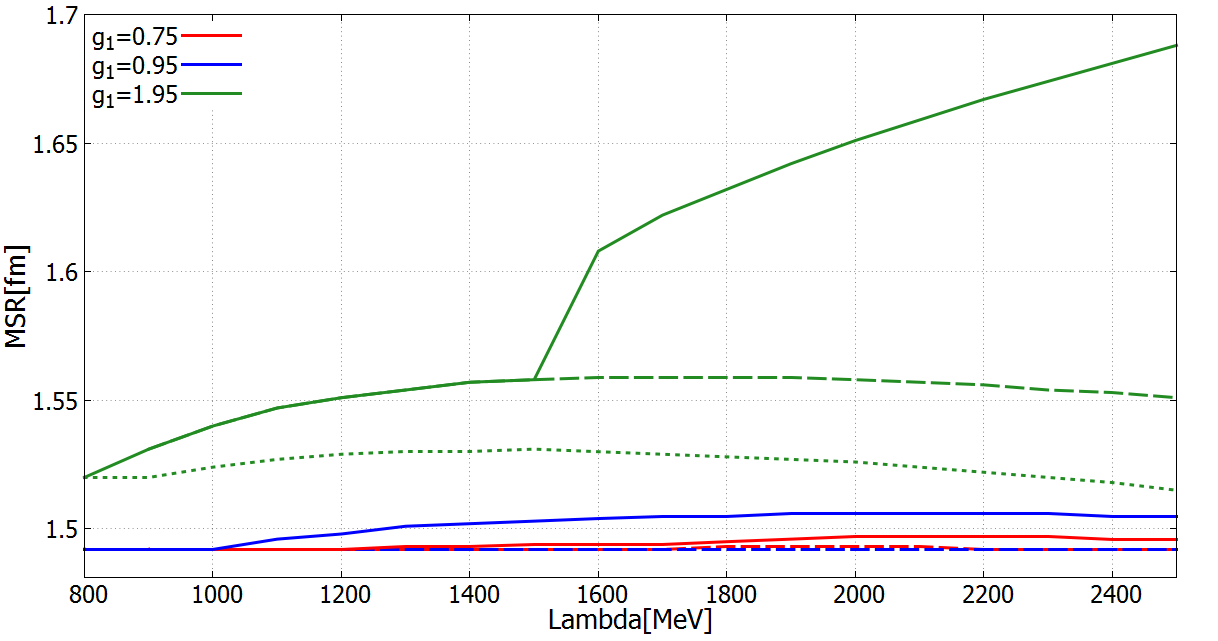} \\
\caption{(color online) (a) Binding energy (B.E.) for the $\Sigma_{b}B^{\ast}$-$\Sigma_{b}^{\ast}B$ molecular state measured from the $\Sigma_{b}^{\ast} B$ threshold of $11113.0$\,MeV, (b) the percentage of the $\Sigma_{b}^{\ast}B$ component and (c) the mean square radius.
The values for the ground states, first excited states and second excited states are shown by solid, dashed and dotted curves, respectively.
The red, blue and green curves are for $g_1 = 0.75$, $0.95$, and $1.95$.
}
\label{fig:b-barb}
\end{center}
\end{figure}

We first study the  molecular state in the coupled channel of 
$\Sigma_{b}$$B^{\ast}$-$\Sigma_{b}^{\ast}$$B$, using
$m_{\Sigma_{b}}=5813.4$\,MeV, $m_{\Sigma_{b}^{\ast}}=5833.6$MeV, $m_{B}=5279.4$\,MeV, $m_{B^{\ast}}=5324.8$\,MeV.
In Fig.~\ref{fig:b-barb}, we show the binding energy measured from the $\Sigma_{b}^{\ast} B$ threshold of $11113.0$\,MeV, together with the percentage of the $\Sigma_{b}^{\ast}B$ component and the mean square radius.
This shows that the values of the binding energy are larger than those for the 
$\Sigma_{c}\bar{D}^{\ast}$-$\Sigma_{c}^{\ast}\bar{D}$ molecular state.
The percentage of the $\Sigma_b^\ast B$ component is slightly smaller for some parameter range, but still more than 99\% in most region.
The value of the mean square radius takes about $1.5$-$1.7$\,fm, some of which are slightly larger than those for the $\Sigma_{c}\bar{D}^{\ast}$-$\Sigma_{c}^{\ast}\bar{D}$ molecular state.
Our results summarized in Fig.~\ref{fig:b-barb} indicate that there exists a hidden bottom pentaquark with mass of about $11080$-$11110$\,MeV and quantum number of $J^P = \frac{3}{2}^-$.
Furthermore, similar to the case for $P_c(4380)$, there may exist two or three molecular states within a few $10$\,MeV range.

\begin{figure}[!htbp]
\begin{center}
(a) \\
\includegraphics[width=0.45\textwidth, bb=0 0 1218 641, clip]{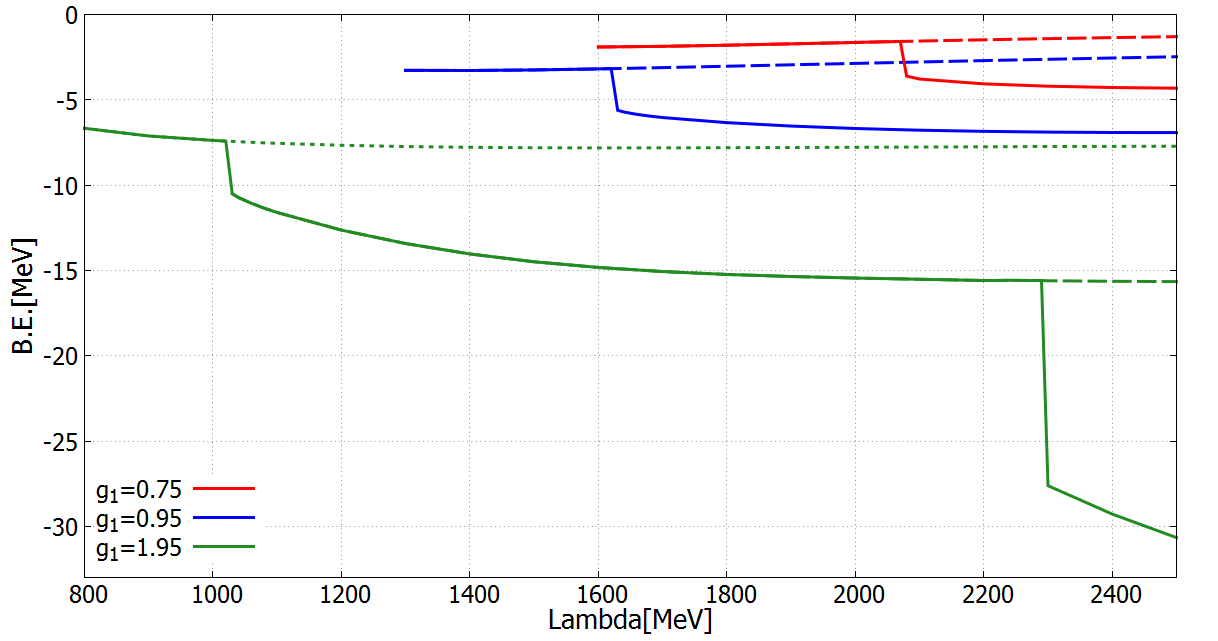} \\
(b) \\
\includegraphics[width=0.45\textwidth, bb=0 0 1218 641, clip]{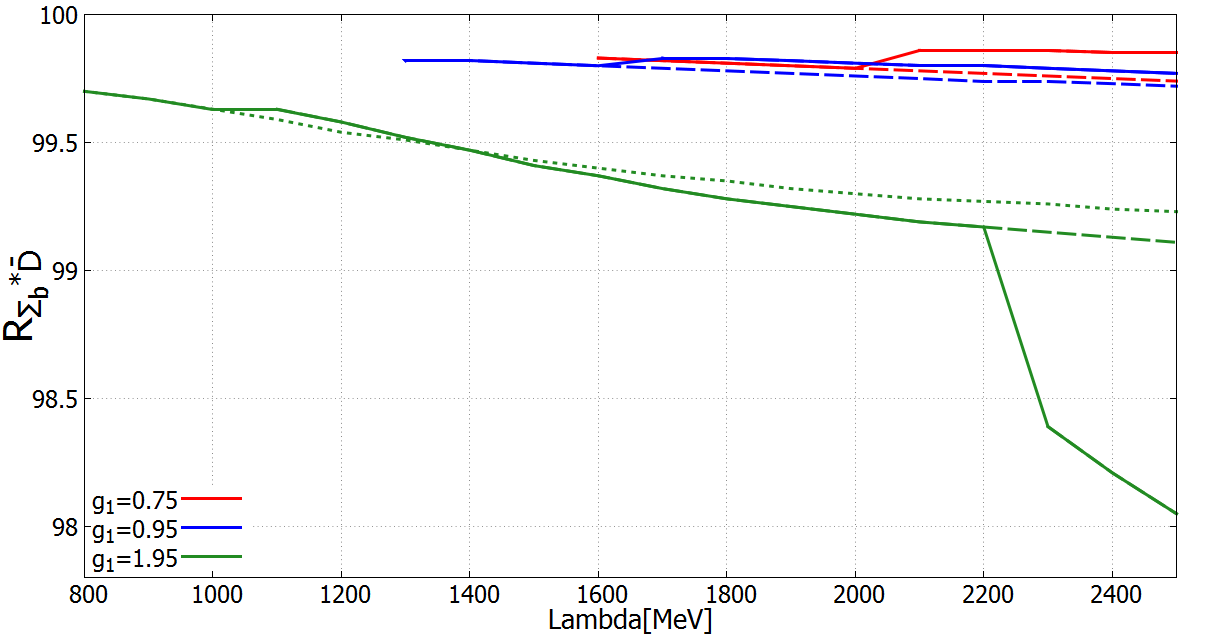} \\
(c) \\
\includegraphics[width=0.45\textwidth, bb=0 0 1218 641, clip]{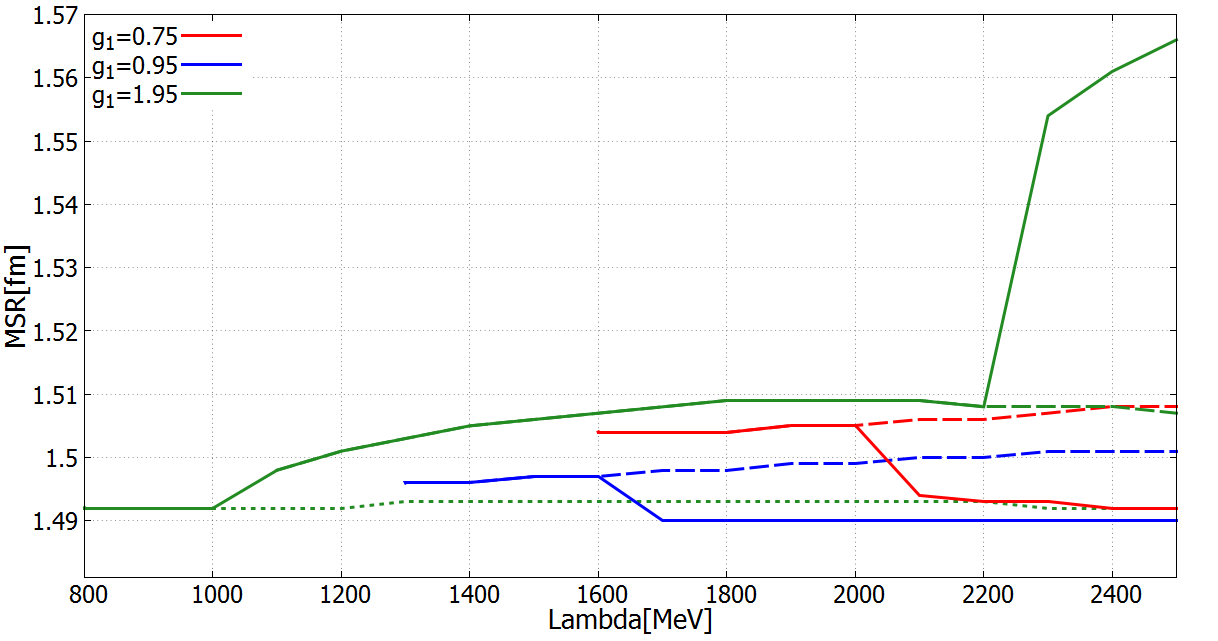} \\
\caption{(color online) 
(a) Binding energy (B.E.) for the $\Sigma_{b}\bar{D}^{\ast}$-$\Sigma_{b}^{\ast}\bar{D}$ molecular state measured from the $\Sigma_{b}^{\ast} \bar{D}$ threshold of $7701$\,MeV,  (b) the percentage of the $\Sigma_{b}^{\ast}\bar{D}$ component, and (c) the mean square radius.
The values for the ground states, first excited states, and second excited states are shown by solid, dashed, and dotted curves, respectively.
The red, blue and green curves are for $g_1 = 0.75$, $0.95$, and $1.95$.
}
\label{fig:b-barc}
\end{center}
\end{figure}

\begin{figure}[!htbp]
\begin{center}
(a) \\
\includegraphics[width=0.45\textwidth, bb=0 0 1218 641, clip]{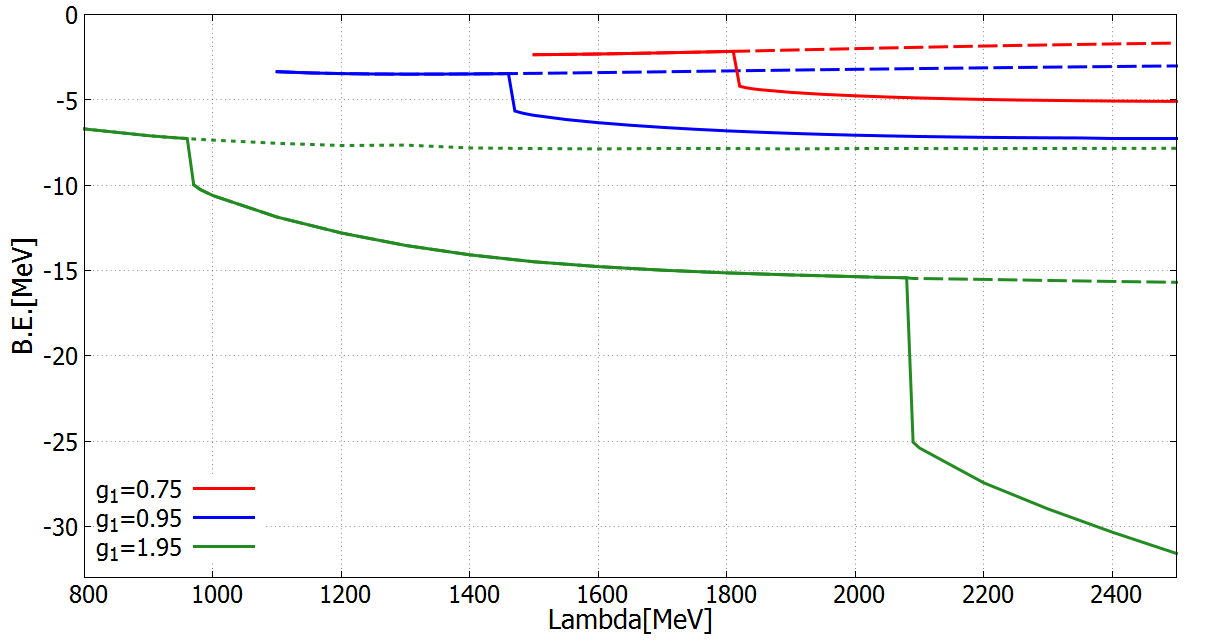} \\
(b) \\
\includegraphics[width=0.45\textwidth, bb=0 0 1218 641, clip]{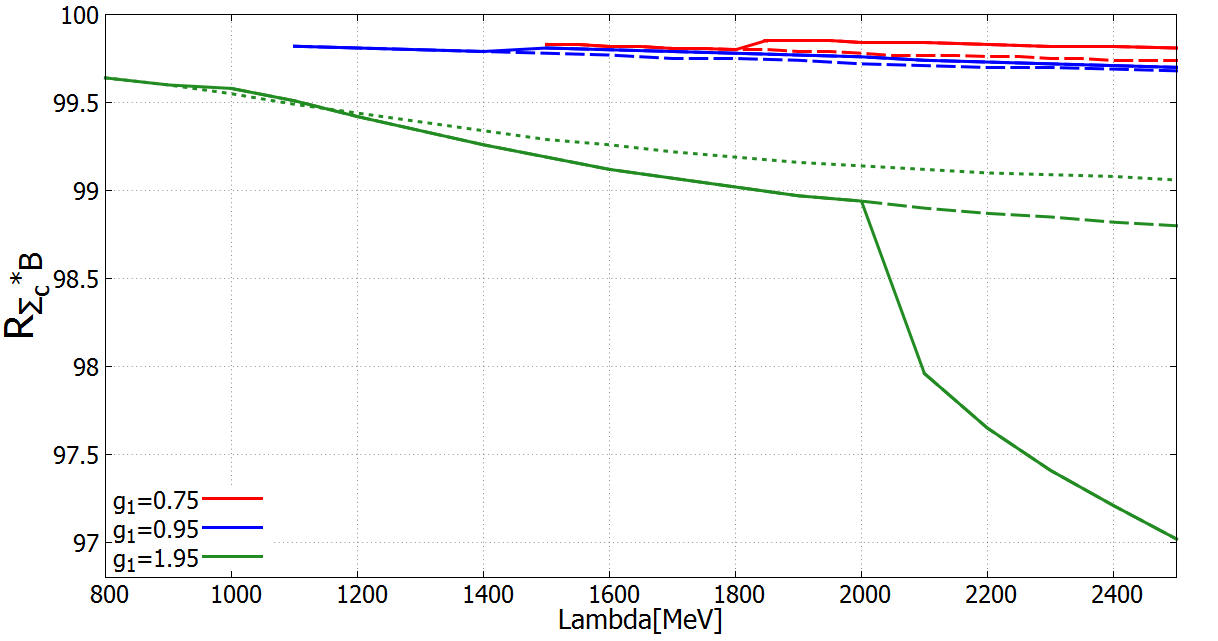} \\
(c) \\
\includegraphics[width=0.45\textwidth, bb=0 0 1218 641, clip]{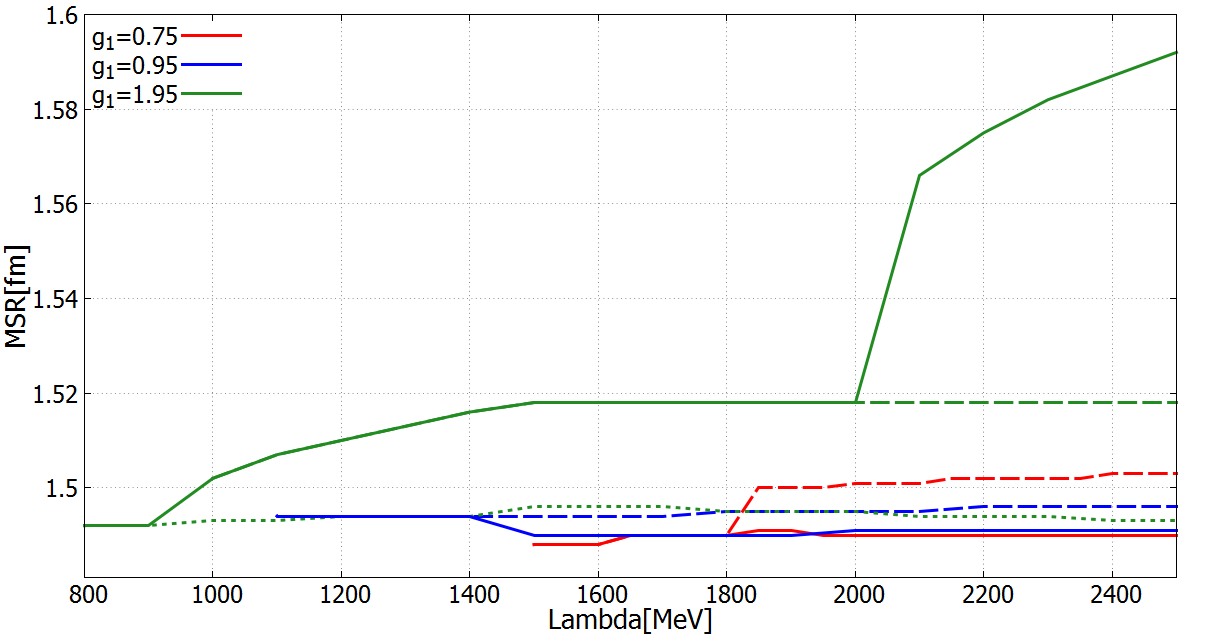} \\
\caption{(color online) 
(a) Binding energy (B.E.) for the $(\Sigma_{c}B^{\ast}, \Sigma_{c}^{\ast}B)$ molecular state measured from the $\Sigma_{c} B^{\ast}$ threshold of $7778$\,MeV,  (b) the percentage of the $\Sigma_{c}^{\ast}B$ component and (c) the mean square radius.
The values for the ground states, first excited states, and second excited states are shown by solid, dashed, and dotted curves, respectively.
The red, blue and green curves are for $g_1 = 0.75$, $0.95$, and $1.95$.
}
\label{fig:c-barb}
\end{center}
\end{figure}

We next study the  molecular states in the coupled channel of 
$\Sigma_{c}$$B^{\ast}$-$\Sigma_{c}^{\ast}$$B$, and that of
$\Sigma_{b}$$\bar{D}^{\ast}$-$\Sigma_{b}^{\ast}$$\bar{D}$, which carry the pure exotic flavor quantum numbers.
In Figs.~\ref{fig:b-barc} and \ref{fig:c-barb}, we show the resultant values of the binding energy, the mixing structure and the mean square radius.
These show that there exist molecular states several MeV below the thresholds, dominantly made from $\Sigma_b^\ast\bar{D}$ or $\Sigma_c^\ast B$, with the size of about $1.5$\,fm.

\begin{figure}[!bp]
\begin{center}
\includegraphics[width=0.45\textwidth, bb=0 0 1218 641, clip]{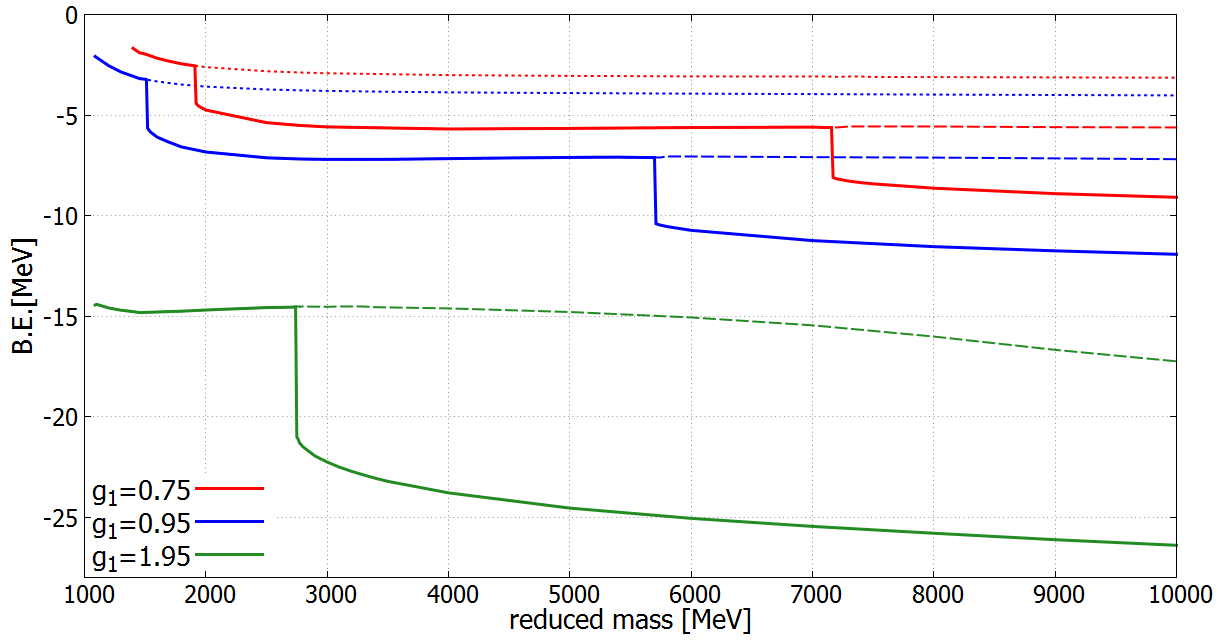}
\caption{(color online) Reduced mass dependence of binding energy (B.E.) with $\Lambda = 1600$\,MeV.
The values of B.E. for the ground states, first excited states and second excited states are shown by solid, dashed and dotted curves.
The red, blue and green curves are for $g_1 = 0.75$, $0.95$, and $1.95$, respectively. 
}
\label{fig:L=1600}
\end{center}
\end{figure}

The results for the binding energy in Figs.~\ref{fig:b-barb}-\ref{fig:c-barb} combined with those in Fig.~\ref{fig:c-barc}
indicate that the binding energy is larger 
for the bound state including heavier components. However, the binding energy  
cannot keep growing with increasing reduced mass, since the depth of the potential is fixed by the values of the cutoff $\Lambda$ and the coupling $g_1$. Then, the binding energy is expected to be saturated to a certain value with increasing reduced mass. 
To check this, we show the dependence of the binding energy on the reduced mass with fixed values of the cutoff $\Lambda = 1600$\,MeV in Fig.~\ref{fig:L=1600}.
This shows that the binding energy is actually saturated at a certain value of the reduced  mass. 

\section{A summary and discussions}
\label{sec:Summary}

We investigated the coupled channel effect of $\Sigma_c^\ast$$\bar{D}$-$\Sigma_c$$\bar{D}^\ast$ to the molecular states.
We constructed a one-pion exchange potential 
following the procedure explained in Ref.~\cite{Yamaguchi:2014era}, and solved the Schr\"odinger-type equation of motion.
Our results showed that the binding energy of the ground state is about several MeV below the threshold of $\Sigma_c^\ast\bar{D}$, $4385.3$\,MeV, in wide range of the cutoff $\Lambda$ for the form factor and the unknown coupling constant of $\Sigma_c^\ast$$\Sigma_c$$\pi$.
Furthermore, for some parameter region, there exist two molecular states within a few MeV range.  
This value is quite similar to the one in Ref.~\cite{He:2015cea}, where the attractive force in a single $\Sigma_c^{\ast} \bar{D}$ channel is obtained by the $\sigma$ exchange. 
We would like to stress that, although the one-pion exchange does not provide attractive force in a single $\Sigma_c^{\ast} \bar{D}$ channel, coupled channel effect of $\Sigma_c^{\ast} \bar{D}$ and $\Sigma_c \bar{D}^{\ast}$ makes $\Sigma_c^{\ast}\bar{D}$ bound. 
We also note the value of the binding energy obtained here is smaller compared with the one in a single $\Sigma_c \bar{D}^{\ast}$ channel obtained in Ref.~\cite{Chen:2015loa}. 
This may originate from the difference between our regularization of the potential following Ref.~\cite{Yamaguchi:2014era} and the one in Ref.~\cite{Chen:2015loa}.
We also studied the size and the mixing structure of the molecular states.
We found that the size of the molecule is about $1.5$\,fm and the percentage of the $\Sigma_c^\ast$$\bar{D}$ component is more than 99\%.
These results indicate that the observed $P_c(4380)$ can be reasonably understood as a loosely bound molecular state dominantly made from the $\Sigma_c^\ast$ baryon and the $\bar{D}$ meson.
We would like to stress that the $\Sigma_c^\ast$ baryon and the $\bar{D}$ meson can form a molecular state mediated by one-pion exchange because the coupled channel effects are included.

We further extended our analysis to the pentaquarks including a $b$ quark and/or an anti-$b$ quark.
Our results showed that there exists a loosely bound molecular state dominantly made from one of the ($\Sigma_c^\ast$, $\Sigma_b^\ast$) baryons and one of the ($\bar{D}$, $B$) mesons, and that the size is always about $1.5$\,fm.
We expect that the existence of these pentaquarks will be tested in future experiments.

In the present analysis, we focus on the $S$-wave bound states, and we do not include the effects of the tensor force by the one-pion exchange.
We expect that inclusion of the tensor force by considering the mixing to the $D$-wave states 
makes the binding energy larger.
In addition, inclusion of other channels may modify the properties of the bound states. 

The present analysis can be extended to the $P$-wave and $F$-wave state of the $\Sigma_c^*\bar{D}$-$\Sigma_c\bar{D}^*$ channel which can be expected to give some explanations of the recently observed $P_c(4450)$. In this case, $P_c(4450)$ can be regarded as the Feshbach resonance state since the mass of $P_c(4450)$ is greater than  the value of the $\Sigma_c^*\bar{D}$ threshold and smaller than that of the $\Sigma_c\bar{D}^*$ threshold.

It will be also very interesting to study the decays of the molecular states obtained in this analysis. 
One possible way is to apply the complex scaling method adopted in, e.g., Ref.\cite{Aoyama:2006CSM}.

We leave the above analyses for future publications.

\acknowledgments
The authors would like to thank Shigehiro Yasui for useful discussions. This work was supported in part by the JSPS Grant-in-Aid for Scientific Research (C) No.~24540266 and No.~16K05345.



\end{document}